\documentclass{article}
\usepackage{main,times}

\usepackage[colorlinks=true, citecolor=green!50!black, linkcolor=blue]{hyperref}

\usepackage{amsmath,amsfonts,bm}

\def\eqref#1{equation~\ref{#1}}

\def\1{\bm{1}}

\DeclareMathAlphabet{\mathsfit}{\encodingdefault}{\sfdefault}{m}{sl}
\SetMathAlphabet{\mathsfit}{bold}{\encodingdefault}{\sfdefault}{bx}{n}

\usepackage{url}
\usepackage{graphicx}
\usepackage{tikz}
\usepackage{subcaption}
\usepackage[most]{tcolorbox}
\title{Vaiage: A Multi-Agent Solution to Personalized Travel Planning}

\author{
\begin{minipage}[t]{0.31\textwidth}
\centering
\textbf{Jiexi Ge}\thanks{Correspondence: liubinwen@stu.xjtu.edu.cn. These authors contributed equally.\newline Our code is available at: \href{https://github.com/Binwen6/Vaiage}{\url{https://github.com/Binwen6/Vaiage}} \newline Our video presentation and slides are available at: \href{https://drive.google.com/drive/folders/1jpp6yo29zyilfbyRx9H7vkYk2xcGPWcx?usp=sharing}{\url{https://drive.google.com/drive/folders/1jpp6yo29zyilfbyRx9H7vkYk2xcGPWcx?usp=sharin}}}\\
3040873839\\
jessie-0918@berkeley.edu
\end{minipage}
\hfill
\begin{minipage}[t]{0.31\textwidth}
\centering
\textbf{Binwen Liu}\footnotemark[1] \\
3040938839\\
liubinwen@berkeley.edu
\end{minipage}
\hfill
\begin{minipage}[t]{0.31\textwidth}
\centering
\textbf{Jiamin Wang}\footnotemark[1] \\
3040938254\\
post1ude@berkeley.edu 
\end{minipage}
}

\newcommand{\VaiageGitHub}{\href{https://github.com/Binwen6/Vaiage}{\url{https://github.com/Binwen6/Vaiage}}}

\iclrfinalcopy
\begin{document}

\maketitle

\begin{abstract}
Planning trips is a cognitively intensive task involving conflicting user preferences, dynamic external information, and multi-step temporal-spatial optimization. Traditional platforms often fall short—they provide static results, lack contextual adaptation, and fail to support real-time interaction or intent refinement.

Our approach, \textbf{Vaiage}, addresses these challenges through a graph-structured multi-agent framework built around large language models (LLMs) that serve as both \textit{goal-conditioned recommenders} and \textit{sequential planners}. LLMs infer user intent, suggest personalized destinations and activities, and synthesize itineraries that align with contextual constraints such as budget, timing, group size, and weather. Through natural language interaction, structured tool use, and map-based feedback loops, Vaiage enables adaptive, explainable, and end-to-end travel planning grounded in both symbolic reasoning and conversational understanding.

To evaluate Vaiage, we conducted human-in-the-loop experiments using rubric-based GPT-4 assessments and qualitative feedback. The full system achieved an average score of \textbf{8.5/10}, outperforming the no-strategy (7.2) and no-external-API (6.8) variants, particularly in \textit{Feasibility}. Qualitative analysis indicated that agent coordination—especially the Strategy and Information Agents—significantly improved itinerary quality by optimizing time use and integrating real-time context. These results demonstrate the effectiveness of combining LLM reasoning with symbolic agent coordination in open-ended, real-world planning tasks.

\end{abstract}

\section{Introduction}

Planning a trip involves far more than selecting a few destinations—it requires balancing diverse
user preferences, navigating real-time constraints, and making coherent decisions under uncertainty.
Yet, most travel planning tools available today offer static results, ignore real-world conditions, and
provide little room for dynamic user interaction or adaptation.

Recent advances in large language models have enabled AI agents to engage in natural, personal-
ized dialogue and reason over open-ended instructions. This opens up new possibilities for travel
planning. However, most existing systems remain limited: recommendation-only platforms such as
iPlan.ai and Itinerary provide little interaction or contextual awareness, while LLM-only assistants
like Skoot hallucinate plausible-sounding but impractical plans due to their lack of grounding in
external data. These approaches fail to deliver adaptive, user-aligned itineraries in realistic travel
scenarios.

We introduce Vaiage, a multi-agent travel planning system that integrates the strengths of LLMs
with real-time information access, modular agent collaboration, and structured decision-making.
Vaiage offers three key capabilities: (1) real-time API integration for environmental awareness (e.g.,
Google Maps, weather, Rapid), (2) adaptive itinerary planning that responds to user goals and con-
straints, and (3) virtual assistant support to refine plans, adjust budgets, and facilitate downstream
actions such as car rental service.

\section{Background}
\label{gen_inst}
\subsection{Temporal and Knowledge-Enhanced Reasoning with LLMs}
Large language models have increasingly demonstrated the ability to perform temporal reasoning within dialogue-based planning tasks. For example, \citet{su2024timo} proposed Timo, which improves a model’s understanding of event sequences by incorporating structured timeline representations. Retrieval-augmented generation methods \citep{lewis2020rag, mao2021retrieval} further enhance planning by allowing models to access external knowledge when reasoning over user queries. To overcome the limitations of short context windows, memory-augmented architectures \citep{khandelwal2019knn, borgeaud2021gopher, wu2022memorizing} enable LLMs to maintain temporal consistency across multiple dialogue turns. In embodied and interactive scenarios, structured prompting has shown effectiveness in multi-step scheduling and decision support \citep{song2023planner, sarch2023embodied}. Building on these advances, Vaiage employs a multi-agent LLM-based system that supports real-time, multi-day itinerary planning by aligning user preferences with time-constrained routes and dynamic travel conditions.

\subsection{Personalized Recommendation with LLMs}
Large language models (LLMs) have increasingly been applied to personalized recommendation tasks by leveraging their capacity for language understanding and adaptive reasoning. \citet{lin2025llmrec} surveyed how LLMs can enhance user modeling and preference prediction within recommender systems. Recent work also explores integrating LLMs with structured retrieval modules for context-aware suggestions, using dialogue-based interaction to refine recommendations over time \citep{li2023dialogrec}, and applying instruction-tuned models for more controllable recommendation outputs \citep{ houlsby2022prompt}. These approaches highlight the role of LLMs in generating flexible, dynamic, and user-aligned content. In our system, a dedicated Recommend Agent processes user preferences and attraction data to produce itinerary-consistent suggestions.

\section{Methodology}
\subsection{External Data Integration}

Vaiage integrates multiple structured APIs to support real-world awareness and personalized planning:

\begin{itemize}
    \item \textbf{Google Maps API}: Used for geocoding, real-time distance and duration estimation, nearby attraction search, and route planning with waypoint optimization.
    
    \item \textbf{OpenWeatherMap}: Provides short-term weather forecasts, enabling weather-aware itinerary adjustments (e.g., indoor attraction prioritization on rainy days).
    
    \item \textbf{Google Places API}: Tourist attractions are fetched and enriched using the Google Places API, combined with heuristics and LLM-based reranking based on user preferences.
    
    \item \textbf{Car Rental API}: Retrieves real-time car rental options through RapidAPI, allowing the system to recommend suitable transportation choices based on trip constraints.
\end{itemize}
\subsection{Multi-Agent Architecture}

Vaiage is built upon a modular multi-agent system in which each agent is responsible for a distinct aspect of the travel planning workflow. Specifically, the system includes:

\begin{figure}[ht]
\centering
\begin{minipage}{0.45\textwidth}
    \centering
    \includegraphics[width=\linewidth]{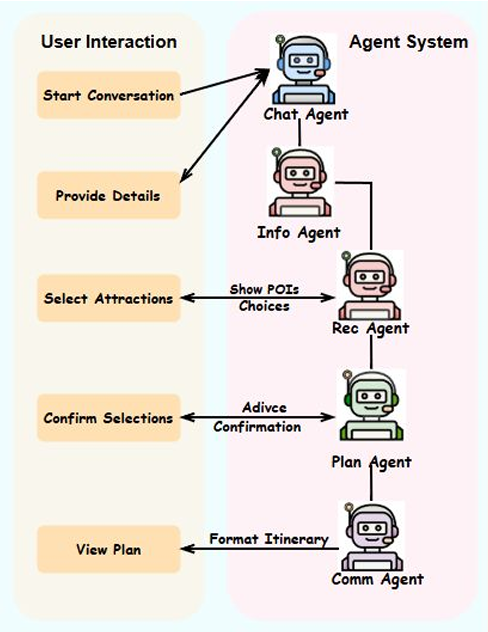} 
    \caption{multi-agent workflow}
    \label{fig:agent-workflow}
\end{minipage}%
\hfill
\begin{minipage}{0.52\textwidth}
    \footnotesize
    \textbf{Chat Agent}: Initiates natural language conversations to extract structured user input, including destination, duration, preferences, and constraints.  \\[4pt]

    \textbf{Information Agent}: Gathers essential real-world information from external APIs such as maps, weather, and rentals. It acts as a shared knowledge provider for downstream agents. \\[4pt]

    \textbf{Recommendation Agent}: Scores and filters attractions using user preferences and LLM-assisted reranking. It ensures diversity, feasibility, and alignment with user-specified needs. \\[4pt]

    \textbf{Route Agent}: Constructs multi-day itineraries and budget. It optimizes for route efficiency, time allocation. \\[4pt]

    \textbf{Strategy Agent}: Refines the plan by analyzing leftover time, adding complementary attractions, and adjusting routes based on duration, constraints, and real-world feasibility. \\[4pt]

    \textbf{Communication Agent}: Produces user-facing messages and manages outputs for potential booking interfaces.
\end{minipage}
\end{figure}

The agents communicate and coordinate through a centralized graph-based context manager, termed \texttt{TravelGraph}. This framework tracks user sessions, maintains persistent state, and enables event-driven message passing between agents. The graph structure allows for reactive updates, making the system robust to incremental changes such as new user constraints or updated API data.

All API calls are handled by the \textbf{Information Agent}, which formats results, caches repeated queries, and integrates them into downstream agents’ planning and decision-making workflows.

\subsection{LLM Reasoning and Prompt Design}

All agents in Vaiage are powered by large language models (LLMs), and are specialized via carefully designed prompt templates. Each agent is aligned to a specific subtask in the planning pipeline, with distinct instructions, input-output schemas, and tool-access configurations.

\textbf{For itinerary planning}, the Route Agent and Strategy Agent employ LLMs to reason over sequences of attractions, travel durations, and time constraints. The Route Agent leverages step-by-step prompting to generate multi-day itineraries that balance duration, location, and daily activity limits, while also estimating costs. The Strategy Agent enhances this by analyzing leftover time and proposing additional attractions that fit both the user’s preferences and practical constraints.

\textbf{For attraction recommendation}, the Recommend Agent uses structured prompts to evaluate user profiles—such as budget, group type, health conditions, and hobbies—against a pool of potential destinations. The LLM re-ranks candidate attractions based on relevance and diversity, considering factors like category balance and weather suitability.

Prompt templates are provided in Appendix~\ref{appendix:prompts}.

\section{User Guides}
Vaiage is an advanced travel planning assistant powered by a sophisticated multi-agent AI architecture, designed to craft personalized, seamless travel experiences. This comprehensive guide provides a detailed walkthrough of Vaiage's features, empowering you to plan your journey with confidence and precision.

\subsection{Getting Started}
\begin{figure}[h]
    \centering
    \includegraphics[width=0.5\textwidth]{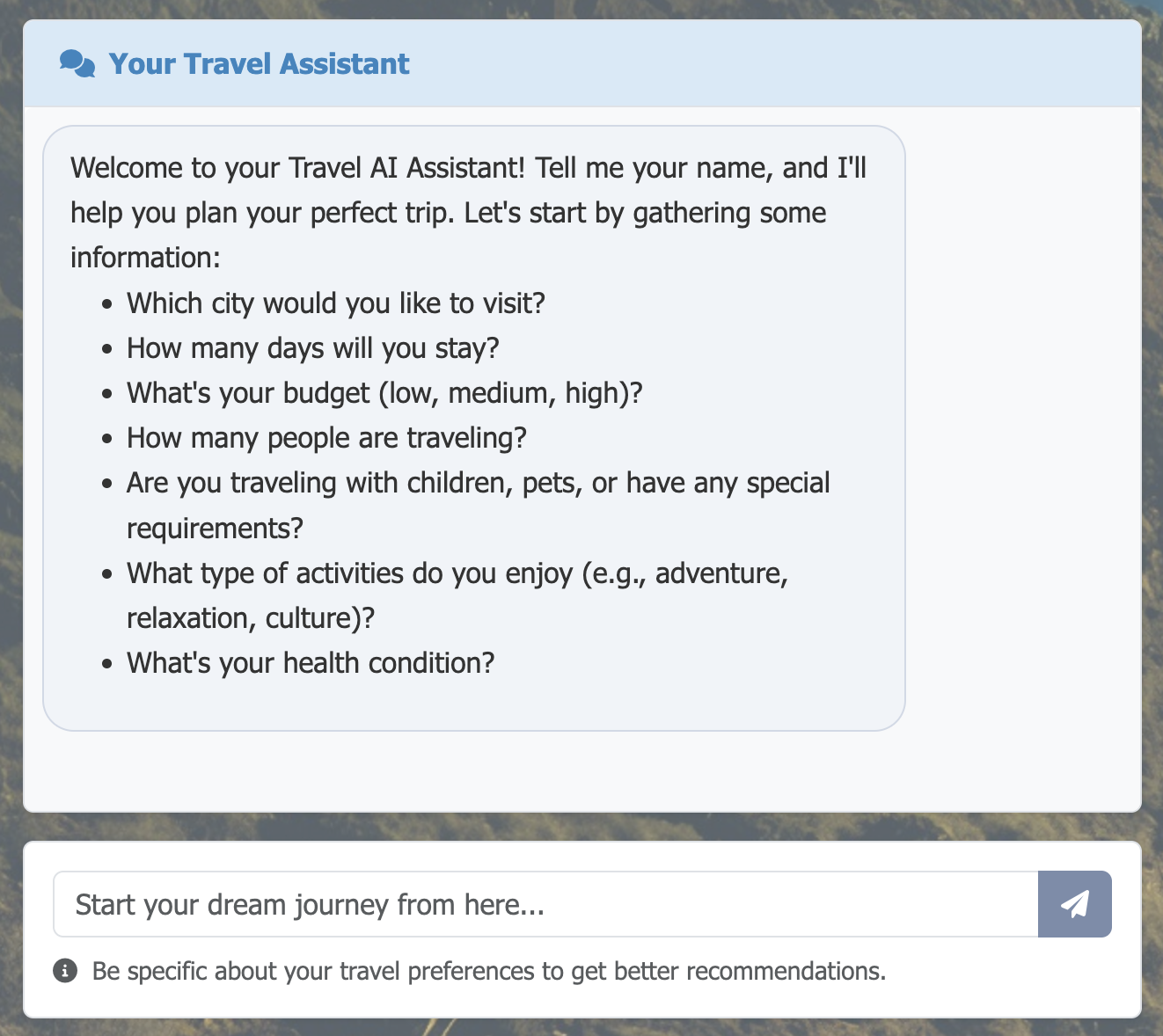}
    \caption{Initial Chat Interface of Vaiage, showcasing the intuitive entry point for travel planning.}
    \label{fig:start}
\end{figure}

Upon launching Vaiage, you are welcomed by an intuitive chat interface \ref{fig:start}, where an AI assistant stands ready to guide you through the travel planning process. Built on a robust multi-agent framework, Vaiage ensures a fluid and tailored experience from the outset, adapting to your preferences and needs.

\subsection{Information Collection and Processing}
\begin{figure}[h]
    \centering
    \includegraphics[width=0.55\textwidth]{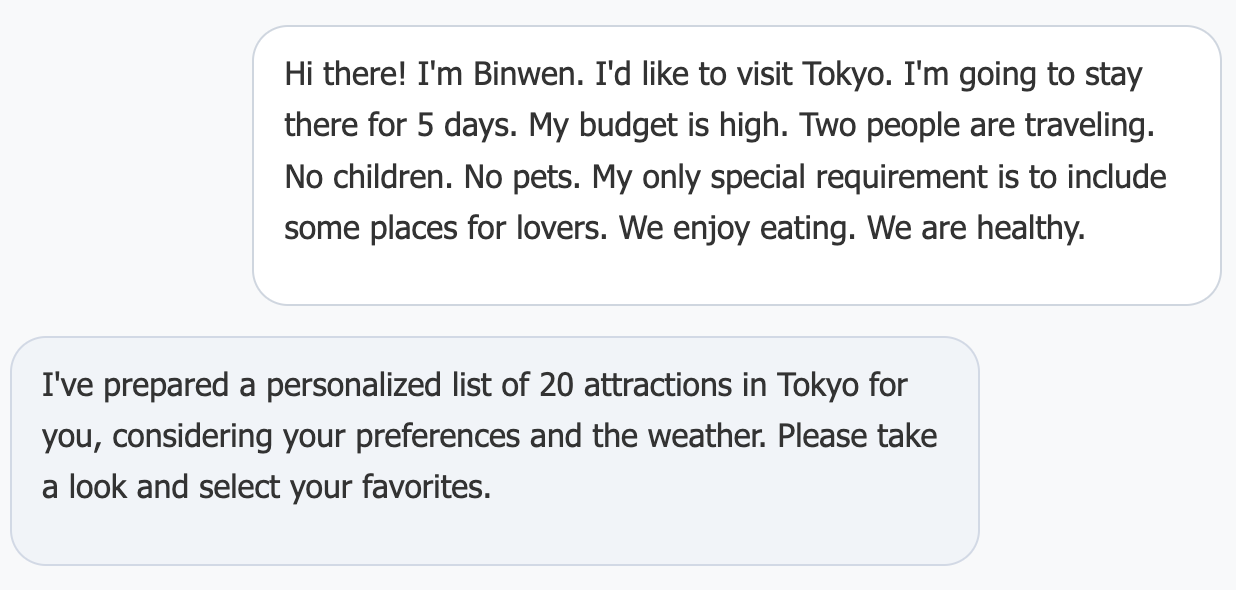}
    \caption{Information Collection and Processing, illustrating how Vaiage transforms user input into actionable travel insights.}
    \label{fig:process}
\end{figure}
Vaiage begins by collecting essential traveler details to craft a personalized itinerary \ref{fig:process}, including names, destination city, travel duration, budget, group size, presence of children, health considerations, hobbies, interests, start date (optional), and special requirements like accessibility or dietary needs. This data is processed intelligently, converting the destination into geographical coordinates, retrieving weather forecasts, and analyzing preferences, local attraction popularity, and accessibility to generate a curated list of attractions tailored to your profile.

\subsection{Attraction Selection}
\begin{figure}[h]
    \centering
    \begin{tikzpicture}
        \node[anchor=north west] (img1) at (0,0.5) {\includegraphics[width=0.25\textwidth]{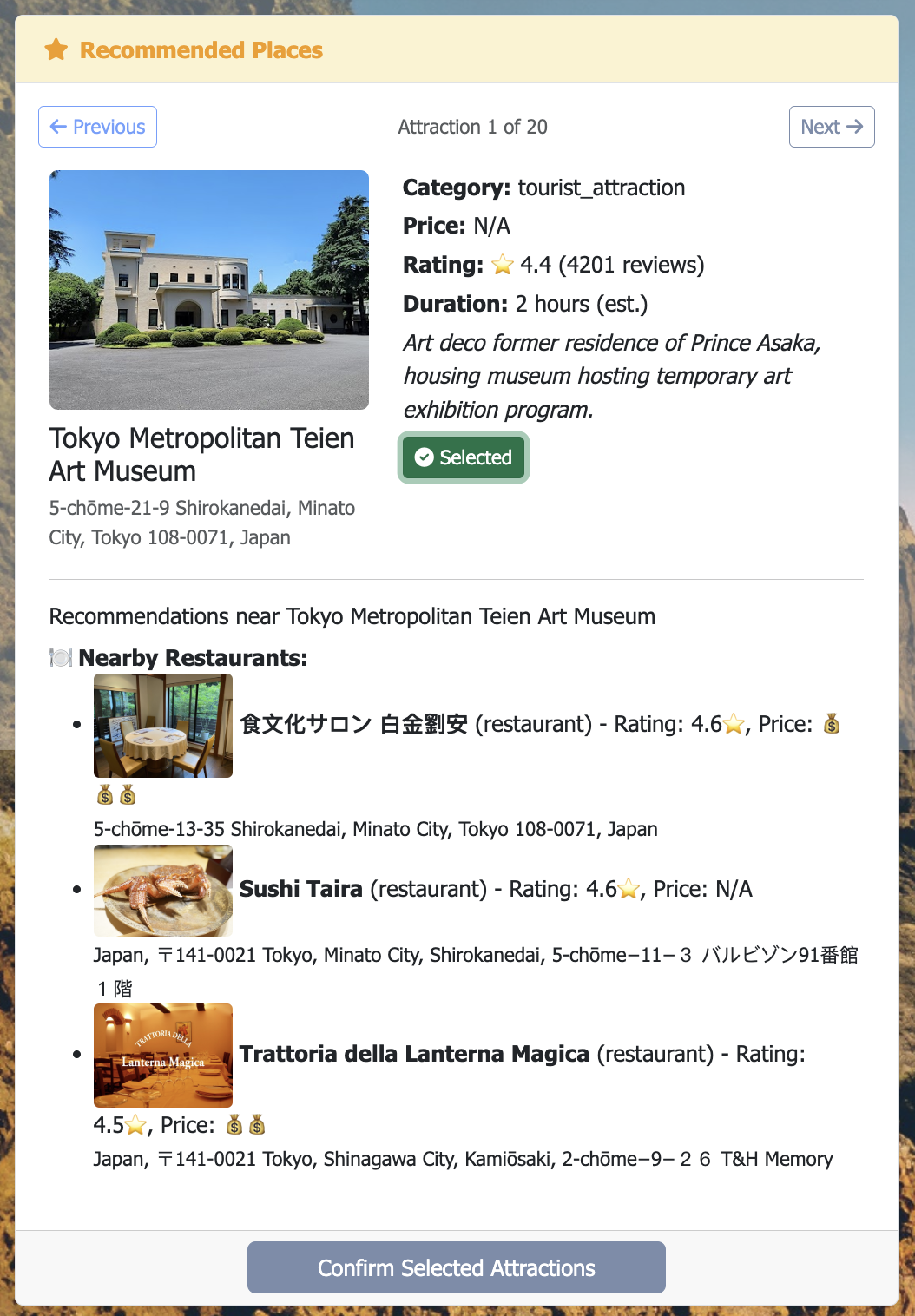}};
        \node[anchor=north west] (img2) at (0.5,0) {\includegraphics[width=0.25\textwidth]{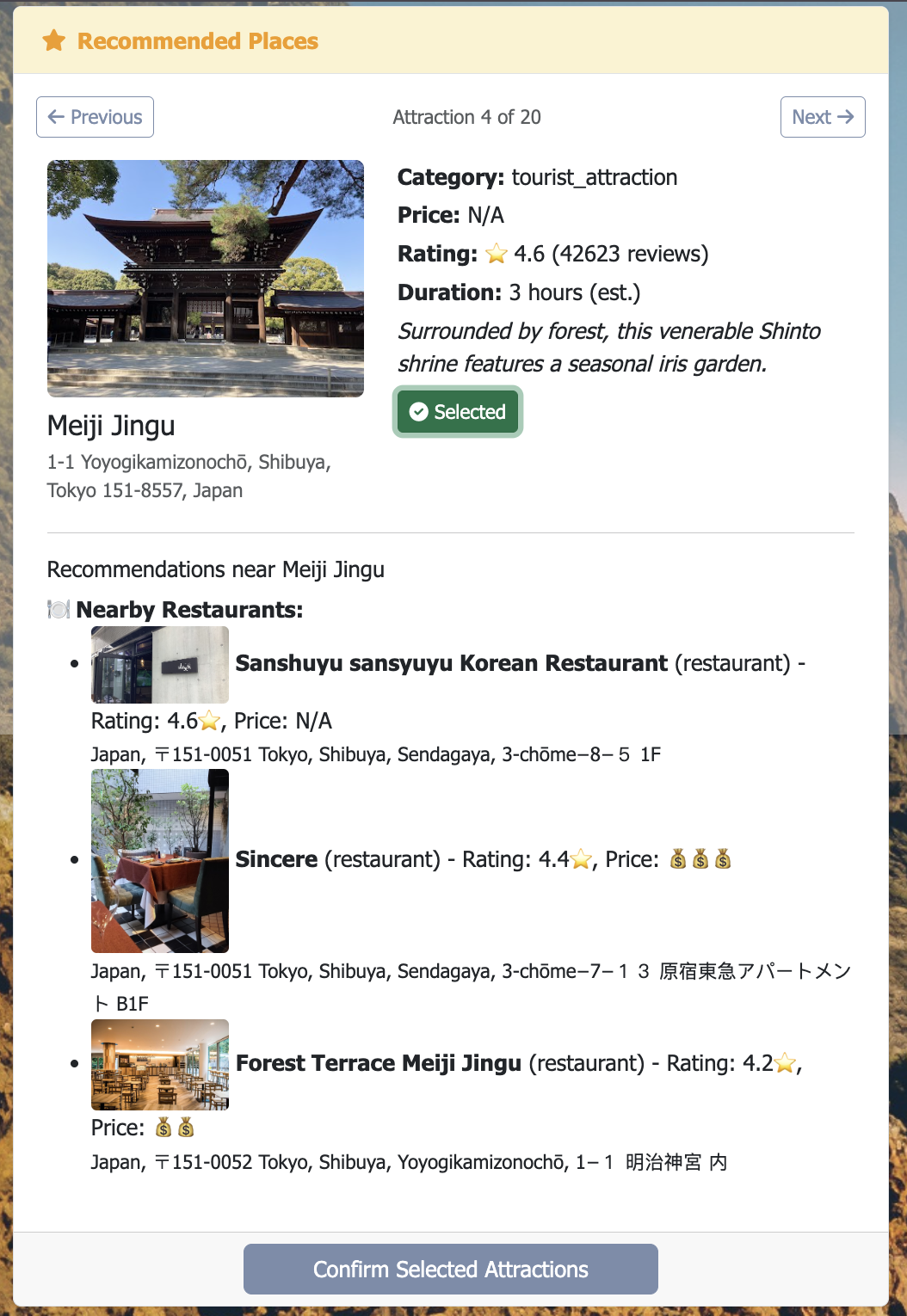}};
        \node[anchor=north west] (img3) at (1.0,-0.5) {\includegraphics[width=0.25\textwidth]{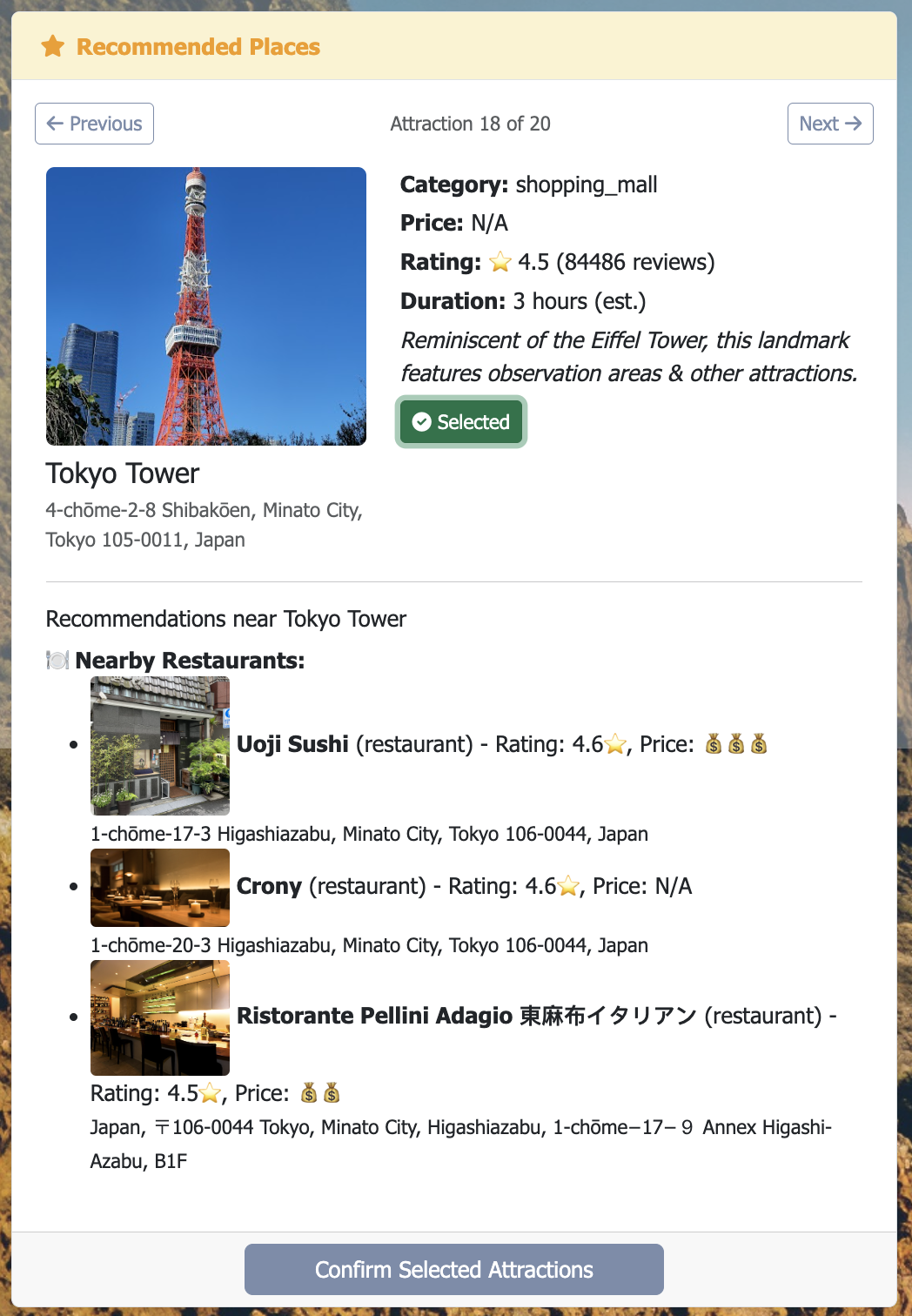}};
        \node[anchor=north west] (img4) at (5,0.5) {\includegraphics[width=0.4\textwidth]{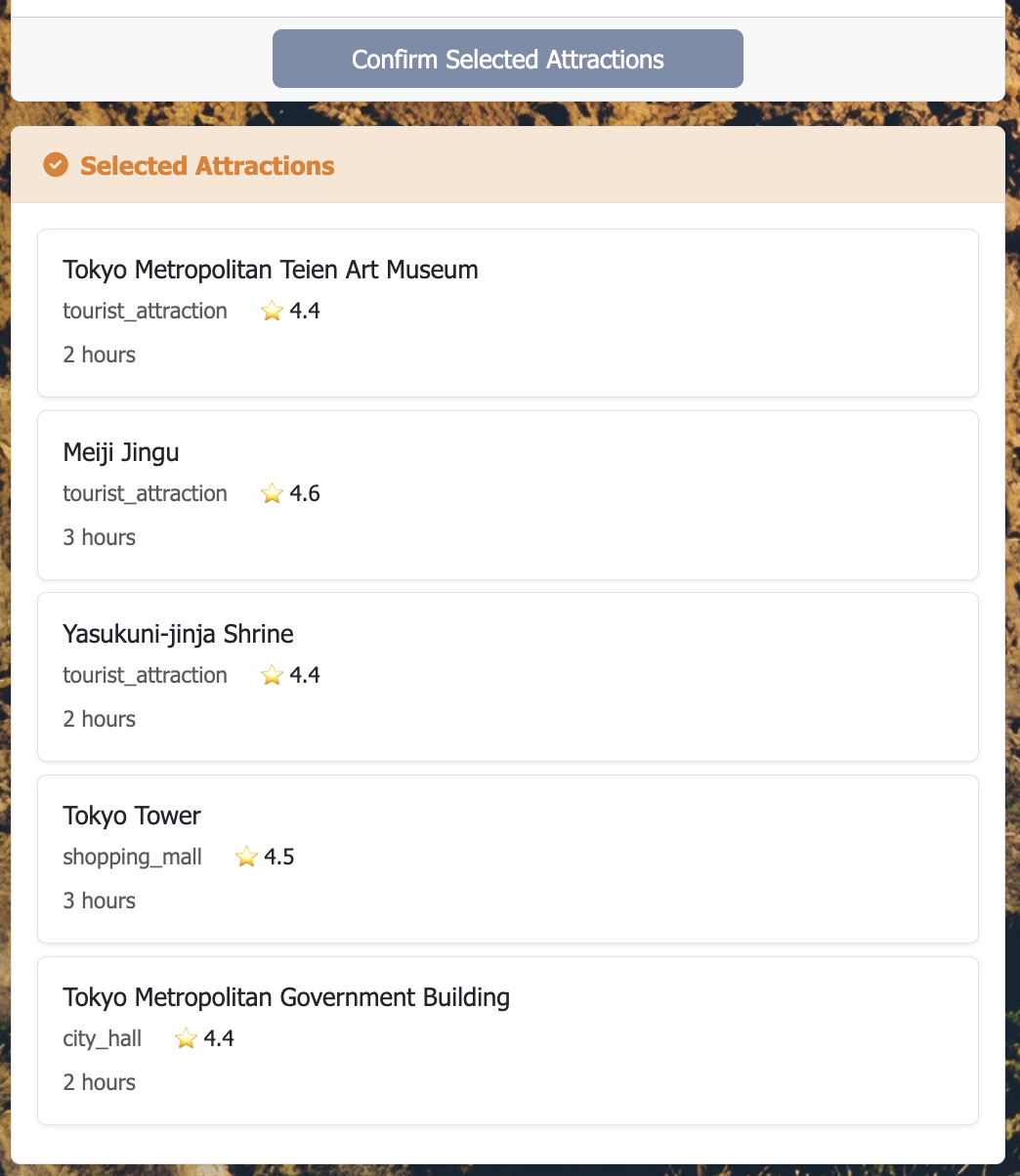}};
    \end{tikzpicture}
    \caption{Attraction Selection Interface: Left – Overlapping options for exploration; Right – Finalized attraction choices.}
    \label{fig:selection}
\end{figure}

With your preferences in mind, Vaiage presents a carefully curated selection of attractions \ref{fig:selection}. The interface allows you to explore detailed information about each attraction, view their locations on an interactive map, and select those that resonate most with your interests. AI-driven recommendations enhance the selection process by suggesting attractions that align with your profile, ensuring a meaningful and engaging travel experience.

\subsection{Strategy Planning}
\begin{figure}[h]
    \centering
    \begin{minipage}[b]{0.48\textwidth}
        \centering
        \includegraphics[width=0.7\textwidth]{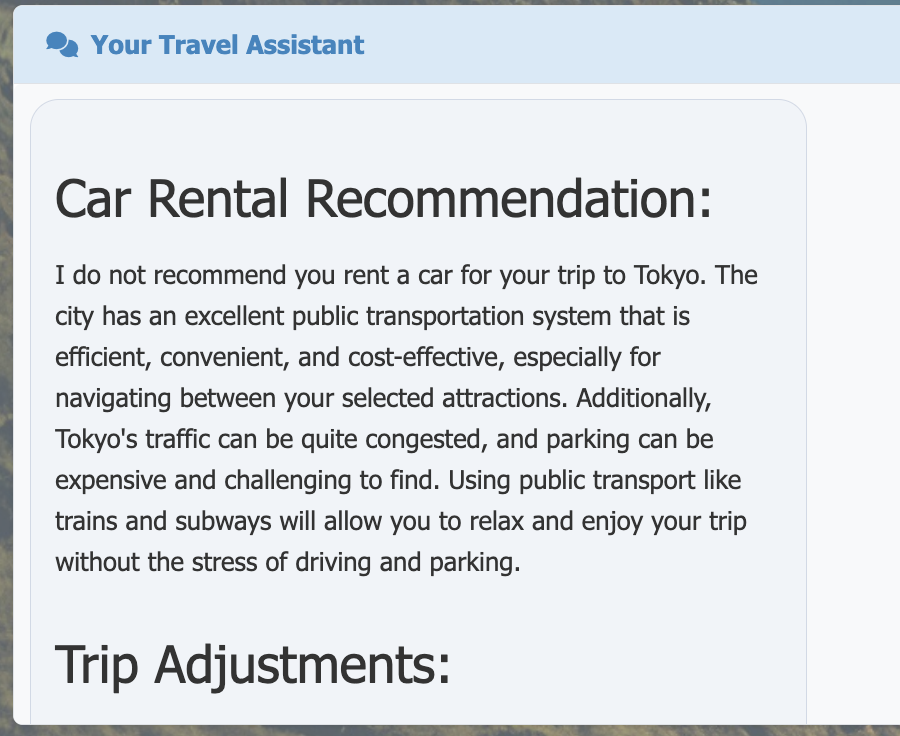}
        \vspace{1em}
        \includegraphics[width=0.7\textwidth]{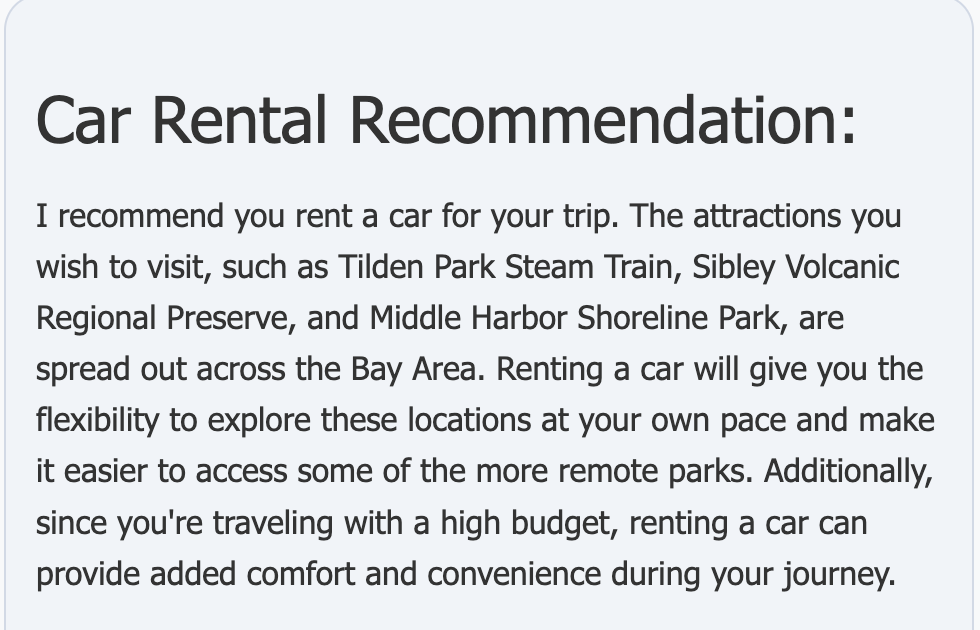}
    \end{minipage}
    \hfill
    \begin{minipage}[b]{0.48\textwidth}
        \centering
        \includegraphics[width=1.1\textwidth]{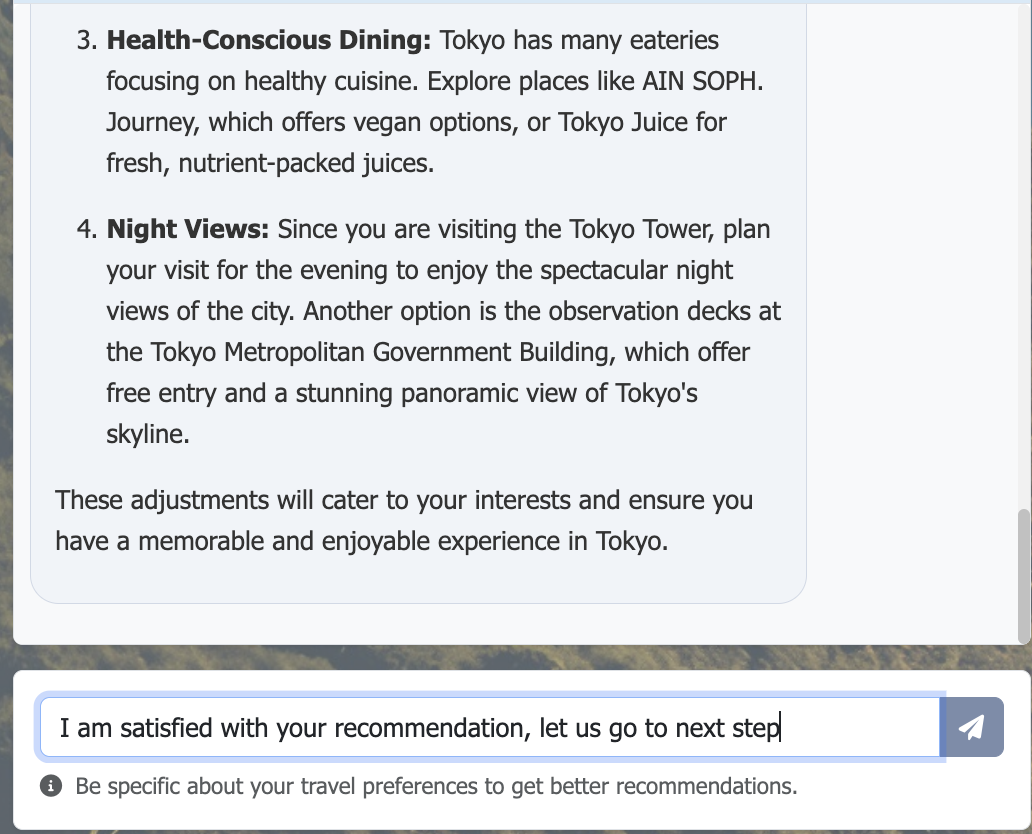}
    \end{minipage}
    \caption{Strategy Planning Interface, highlighting car recommendations and trip adjustments for optimal scheduling.}
    \label{fig:strategy}
\end{figure}

After you select your attractions, Vaiage's strategy planning module takes over to create an optimized itinerary \ref{fig:strategy}. The system analyzes factors such as attraction opening hours, weather conditions, travel distances, and your stated preferences to craft a cohesive visiting plan. Transportation options are recommended to suit your needs, and precise timing suggestions ensure a balanced and enjoyable schedule, maximizing your time at each destination.

\subsection{Route Planning}
\begin{figure}[h]
    \centering
    \includegraphics[width=0.55\textwidth]{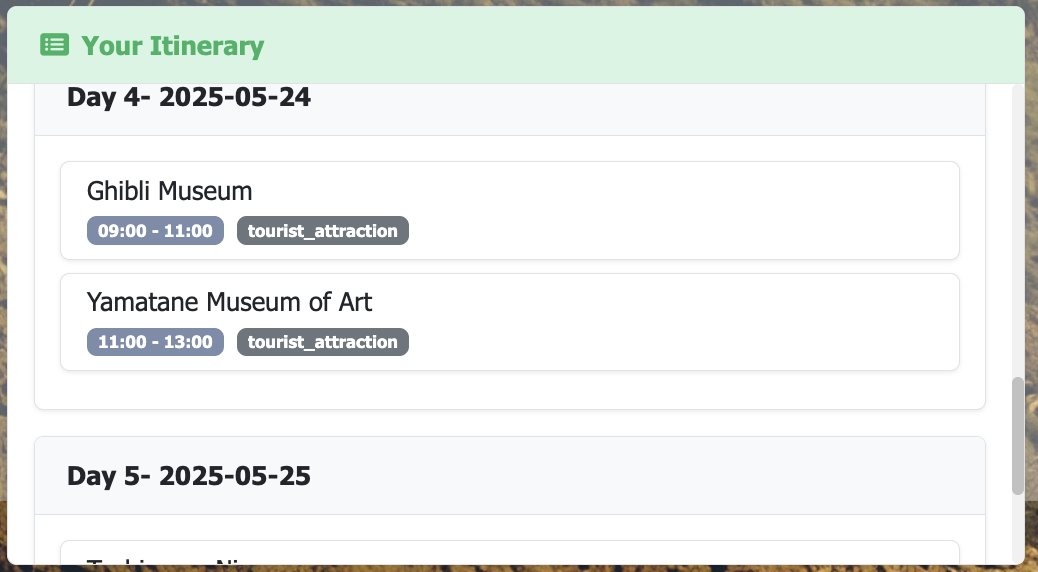}
    \caption{Route Planning Interface, depicting efficient travel paths and transportation options.}
    \label{fig:route}
\end{figure}

Vaiage's route planning feature designs efficient travel paths between your chosen attractions \ref{fig:route}. By considering traffic patterns, public transportation options, walking distances, and time of day, the system generates an optimal visiting order and provides detailed directions. This ensures smooth transitions between locations, allowing you to focus on enjoying your journey rather than navigating logistics.

\subsection{Interactive Map Features}
\begin{figure}[h]
    \centering
    \includegraphics[width=0.7\textwidth]{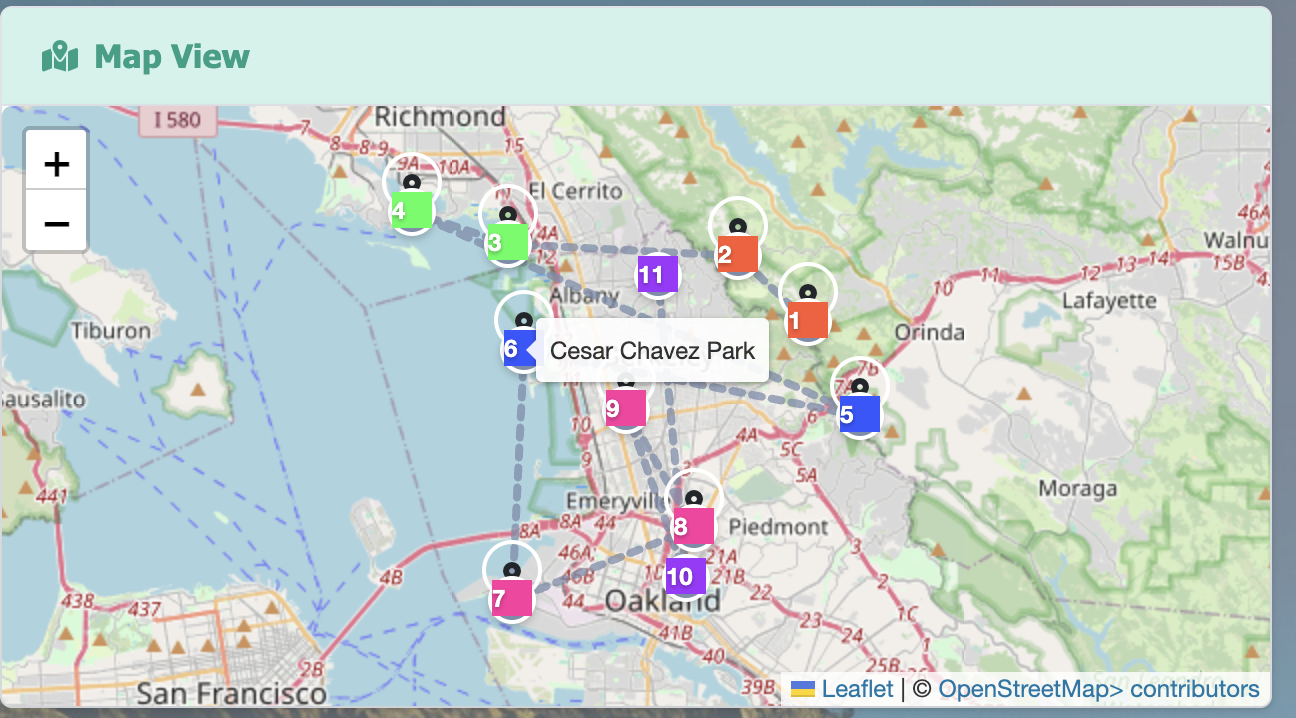}
    \caption{Interactive Map Features, showcasing visual navigation tools and location markers.}
    \label{fig:interact}
\end{figure}

The interactive map is a cornerstone of Vaiage's user experience, offering a dynamic visualization of your itinerary \ref{fig:interact}. Attractions are marked with color-coded icons, and optimal routes are clearly delineated. Nearby amenities, services, and public transportation options are highlighted, while street-level details provide context for your surroundings. This feature empowers you to navigate your destination with confidence and ease.

\subsection{Budget Planning}
\begin{figure}[h]
    \centering
    \includegraphics[width=0.6\textwidth]{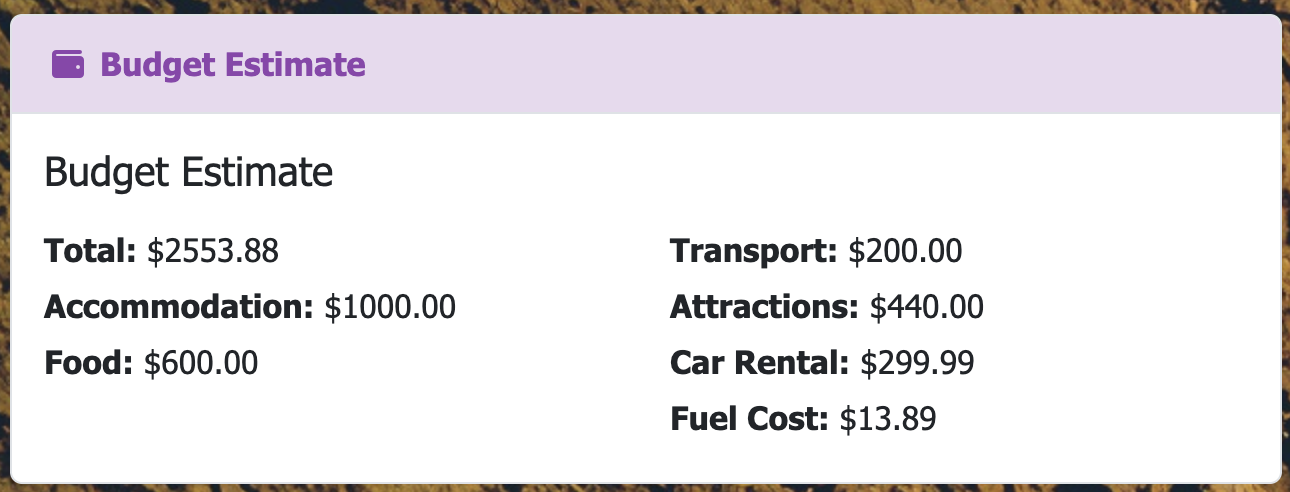}
    \caption{Budget Planning Interface, providing a detailed cost breakdown and budget-friendly suggestions.}
    \label{fig:budget}
\end{figure}

To keep your travel plans financially aligned, Vaiage's budget planning tool estimates costs for attractions, transportation, and other expenses \ref{fig:budget}. A comprehensive cost breakdown is provided, along with suggestions for budget-friendly alternatives where applicable. This transparent approach ensures you can make informed decisions and stay within your financial comfort zone.

\subsection{Trip Confirmation}
\begin{figure}[h]
    \centering
    \begin{subfigure}[t]{0.35\textwidth}
        \centering
        \includegraphics[width=\textwidth]{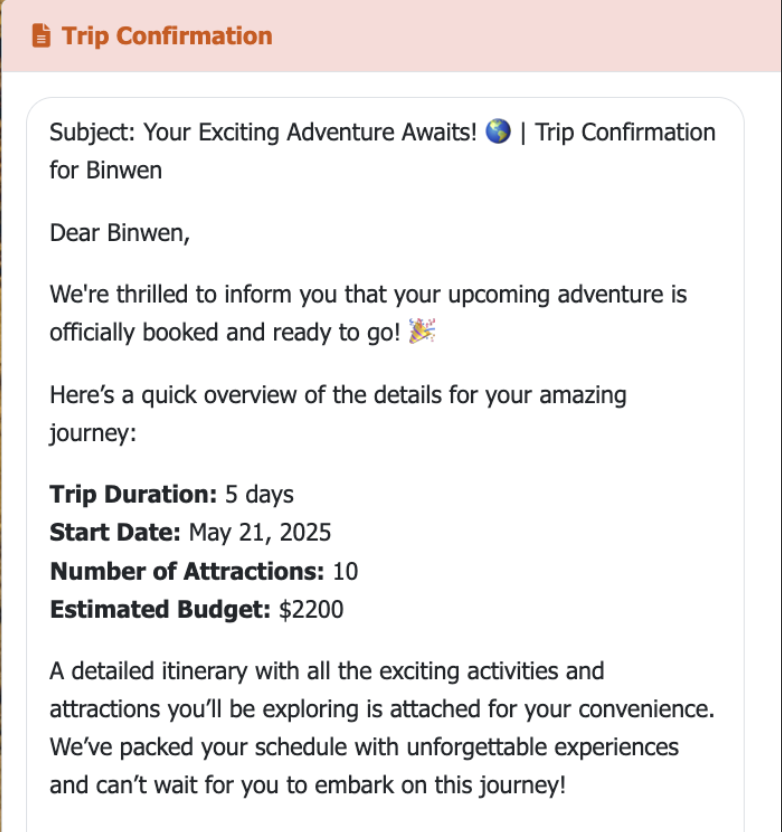}
        \caption{Left side of the Trip Confirmation interface, detailing the itinerary and logistics.}
        \label{fig:trip-left}
    \end{subfigure}
    \hfill
    \begin{subfigure}[t]{0.4\textwidth}
        \centering
        \includegraphics[width=\textwidth]{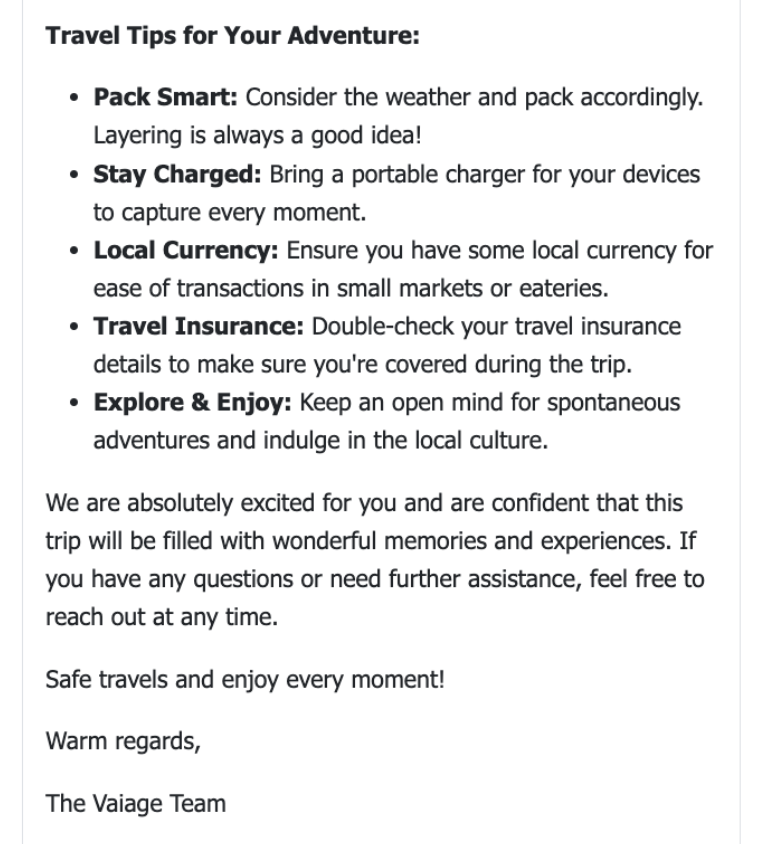}
        \caption{Right side of the Trip Confirmation interface, featuring an interactive map and additional details.}
        \label{fig:trip-right}
    \end{subfigure}
    \caption{The itinerary confirmation interface is divided into itinerary details on the left and an interactive map view on the right.}
    \label{fig:trip}
\end{figure}

The Trip Confirmation phase is the culmination of your planning journey, presenting a comprehensive overview of your itinerary \ref{fig:trip}. The interface is split into two complementary sections. On the left, a detailed itinerary outlines your selected attractions, estimated visiting times, transportation arrangements, weather forecasts, and a full budget breakdown. On the right, an interactive map visualizes your plan, with custom icons marking attractions, optimal routes, color-coded zones, nearby amenities, and public transportation options.

At this stage, you can review and refine your itinerary, adjust visiting times or attraction order, and access real-time weather updates. Vaiage also allows you to export your itinerary in multiple formats or share it with travel companions. A final checklist ensures all critical elements—such as transportation, accommodations, special requirements, and emergency contacts—are in place, guaranteeing a well-organized and stress-free trip.

\subsection{Planning and Optimal Use}
Vaiage automatically saves your planning session, enabling you to revisit, modify, or reset your itinerary at any time for a flexible and convenient experience. To maximize your experience, provide detailed preferences during the information collection phase, engage with AI recommendations by offering feedback, and use the interactive map to visualize your plans. Consider weather forecasts when scheduling activities and review budget estimates to align with your financial goals. Leveraging Vaiage's adaptive learning refines your travel plans to match your evolving preferences.

\subsection{Technical Requirements and Support}
Vaiage requires a modern web browser (e.g., Chrome, Firefox, Safari, or Edge), a stable internet connection, JavaScript enabled, and a recommended screen resolution of at least 1024x768 for optimal accessibility. For support, the chat interface offers immediate AI-driven assistance, and error messages provide specific guidance. To reach further support, users can consult the documentation or contact the development team through the project's GitHub page \VaiageGitHub.

\section{Experiments}

\subsection{Evaluation Setup}

We evaluate Vaiage by assessing the alignment between user-provided trip preferences and the multi-agent system's generated travel plans. Since traditional metrics such as BLEU or ROUGE are not applicable to interactive planning tasks, we adopt a model-based evaluation strategy using large language models (LLMs) as proxy judges. Specifically, we leverage GPT-4 and GPT-3.5 to rate the system outputs along multiple dimensions.

\subsection{Scenarios and Inputs}

We define a diverse set of travel planning scenarios, including:
\begin{itemize}
    \item \textbf{4-day Tokyo cultural tour} with focus on history and cultural sites
    \item \textbf{2-day Los Angeles architecture tour} with mobility considerations
    \item \textbf{3-day San Diego family trip} with children's activities
    \item \textbf{2-day Shanghai shopping and architecture experience}
    \item \textbf{6-day Hong Kong cultural exploration}
\end{itemize}

Each scenario includes:
\begin{itemize}
    \item Natural language user input (e.g. "I am planning a trip with a group of 3 adults to Los Angeles for 4 days. We have a high budget. I am in good health but gets tired easily. We are interested in architecture.")
    \item System-generated response (selected POIs + ordered itinerary + budget allocation)
\end{itemize}

\subsection{LLM-Based Scoring}

We prompt LLMs with the full user request and the system's output, asking them to rate the plan on multiple dimensions:

\begin{itemize}
    \item \textbf{Relevance}: Do the chosen POIs align with the user's preferences and interests?
    \item \textbf{Feasibility}: Is the itinerary logically and geographically coherent? Does it consider practical constraints like opening hours and travel time?
    \item \textbf{Personalization}: Does the plan account for user-specific factors like health conditions, budget constraints, and group composition?
    \item \textbf{Satisfaction}: Would a typical user likely accept this plan as-is? Does it provide a good balance of activities?
\end{itemize}

Scoring is done on a 1-10 scale, with detailed justifications for each rating. We compare results across different variants:
\begin{itemize}
    \item Full system (all agents)
    \item Ablated: without the Strategy Agent
    \item Ablated: without external APIs (weather, POI data)
\end{itemize}

\subsection{Results and Findings}

Results show that the full system significantly outperforms ablated baselines, especially in terms of Relevance and Personalization. GPT-4 ratings for the full system averaged 8.5/10 across all scenarios, while the no-Strategy variant dropped to 7.2. The no-external-APIs variant showed the most significant drop in Feasibility scores (6.8/10), highlighting the importance of real-time data integration.

Key findings include:
\begin{itemize}
    \item The system demonstrates strong capability in handling diverse user preferences, with particularly high scores (9/10) for architecture-focused and cultural tours
    \item Weather-aware planning contributed notably to Feasibility scores, especially for outdoor activities
    \item Budget allocation shows good alignment with user-specified constraints, with appropriate adjustments for group size and activity types
    \item The system effectively balances multiple competing factors (e.g., health conditions, group preferences, time constraints)
\end{itemize}

Qualitative analysis indicates that agent coordination improves contextual alignment, especially when combining POI filtering with user preference memory. The Strategy Agent's role in optimizing daily schedules and the Information Agent's contribution to gathering real-time data were particularly crucial for high-quality plan generation.

Sample plan ratings and output snapshots are presented in Appendix~\ref{appendix:evaluation}. The evaluation data, including detailed LLM assessments and human evaluations, is available in the project repository.

\section{Conclusion}

In this work, we introduce \textbf{Vaiage}, a modular multi-agent system for personalized travel planning. By combining LLMs, real-time data, and structured coordination, Vaiage delivers dynamic, user-adaptive itineraries beyond static tools.

Its agent specialization and graph memory enable flexible, goal-driven collaboration. Future work includes long-term personalization, collaborative planning, and booking integration. Vaiage highlights the potential of LLM-based agents in real-world decision-making tasks.

\bibliographystyle{plainnat}  
\bibliography{references}   
\nocite{*}

\newpage
\appendix

\section{Prompt Templates}
\label{appendix:prompts}

We show below the core prompt templates used by selected agents. These prompts are used to guide the behavior of LLMs in planning and recommendation.

\subsection{Strategy Agent Prompt (Itinerary Generation)}

\begin{tcolorbox}[
    title=Prompt Template – Strategy Agent,
    fonttitle=\bfseries,
    colback=white,
    colframe=black,
    coltitle=white,
    colbacktitle=black,
    sharp corners,
    boxrule=0.5pt,
    width=\textwidth,
    enhanced,
    breakable
]

\textbf{You are a travel advisor helping with trip logistics. The user is planning a \{N\}-day trip.}

\vspace{0.5em}

\textbf{User Preferences:} Structured input including hobbies, health conditions, mobility status, group composition (e.g., with children), and budget level.

\textbf{Weather Summary:} Forecasted conditions during the trip (e.g., rainy, sunny).

\textbf{Pre-selected Attractions:} A list of must-include attractions that must appear in the final plan.

\textbf{Available Attractions:} A full list of candidate attractions (with durations and locations), possibly overlapping with pre-selected ones.

\vspace{0.5em}
\textbf{Instructions:}
\begin{itemize}
    \item Distribute attractions across the \{N\} days assuming 8 hours per day.
    \item Prioritize proximity to minimize travel time.
    \item Adapt scheduling based on:
    \begin{itemize}
        \item Health or mobility constraints (e.g., fewer spots per day)
        \item Family-friendly needs (e.g., include child-oriented content)
        \item User interests (e.g., history, nature, food)
        \item Budget level (e.g., avoid high-cost attractions)
        \item Weather conditions (e.g., prefer indoor spots on rainy days)
    \end{itemize}
\end{itemize}

\vspace{0.5em}
\textbf{Output Format:}  
Return a valid JSON object mapping each day to a list of attraction names.\\
Example: 
\begin{verbatim}
{"day1": ["Attraction A", "Attraction B"],
 "day2": ["Attraction C", "Attraction D"]}
\end{verbatim}
Do not include any markdown or bold formatting.
\end{tcolorbox}

\subsection{Recommend Agent Prompt (Attraction Reranking)}

\begin{tcolorbox}[
    title=Prompt Template – Recommend Agent,
    fonttitle=\bfseries,
    colback=white,
    colframe=black,
    coltitle=white,
    colbacktitle=black,
    sharp corners,
    boxrule=0.5pt,
    width=\textwidth,
    enhanced,
    breakable
]
You are an expert travel recommender. Your task is to rank the provided list of attractions based on the user's preferences, the details of each attraction, and the current weather summary.

\vspace{0.5em}
\textbf{User Preferences:} Structured profile including hobbies, health conditions, mobility status, group composition (e.g., with children), and budget level.

\textbf{Weather Summary:} Forecasted conditions during the trip period.

\textbf{Attractions List:} Each attraction includes metadata such as \texttt{id}, \texttt{name}, \texttt{category}, \texttt{estimated\_duration}, \texttt{price\_level}, \texttt{rating}, and a short \texttt{description} (if available).

\vspace{0.5em}
\textbf{Ranking Criteria:}
\begin{itemize}
    \item \textbf{User Hobbies \& Interests:} Match to stated hobbies (e.g., history, art, nature).
    \item \textbf{Health \& Accessibility:} Adjust for physical limitations or low endurance.
    \item \textbf{Child Suitability:} If traveling with children, prioritize family-friendly options.
    \item \textbf{Budget:} Respect financial constraints and filter out high-cost spots when needed.
    \item \textbf{Weather Compatibility:} Match indoor/outdoor types with weather forecast.
    \item \textbf{Category Balance:} Ensure diversity in top suggestions; remove duplicates.
\end{itemize}

\vspace{0.5em}
\textbf{Output Format:}  
Return a valid JSON list of attraction \texttt{id}s ranked from most to least recommended.\\
Example:
\begin{verbatim}
["id1", "id2", "id3"]
\end{verbatim}
Do not include any explanation or extra formatting.
\end{tcolorbox}

\newpage

\section{Evaluation}
\label{appendix:evaluation}

\subsection{Evaluation Scores}
\begin{figure}[h]
    \centering
    \includegraphics[width=0.75\textwidth]{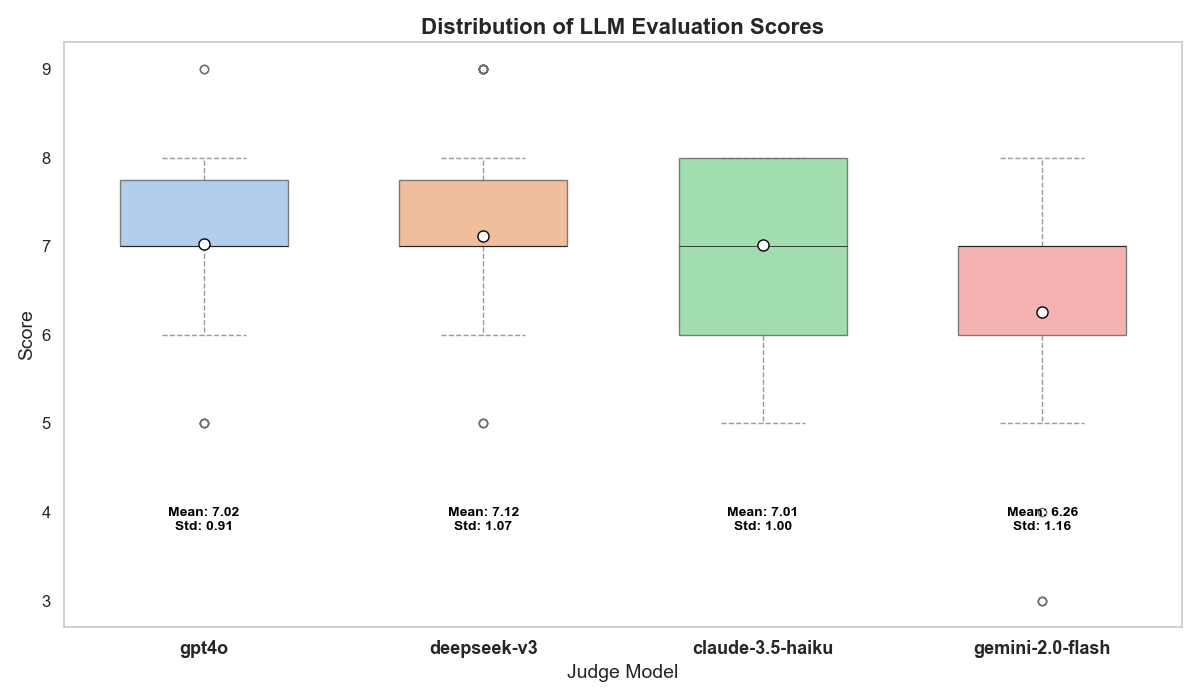}
    \caption{Distribution of LLM Evaluation Scores}
    \label{fig:boxplot}
\end{figure}

\subsection{Output Samples}

\subsubsection{Architecture-Focused Tour (High Score Example)}
\label{appendix:architecture}

\textbf{User Request:} "Hello, I'm Emma Wilson. I'm planning a trip with a group of 3 adults to Los Angeles for 4 days. We haven't decided on a start date yet. We have a high budget. I am in good health but gets tired easily. We are interested in architecture."

\textbf{System Response:}
\begin{itemize}
    \item \textbf{Agent Recommendation:} Car rental is recommended
    \item \textbf{Selected Attractions:}
    \begin{itemize}
        \item Walt Disney Concert Hall (Frank Gehry-designed music hall)
        \item Los Angeles City Hall (1928 landmark)
        \item Bradbury Building (Victorian architecture)
        \item Natural History Museum
        \item California Science Center
    \end{itemize}
    \item \textbf{Budget Allocation:}
    \begin{itemize}
        \item Total: \$2,901.96
        \item Accommodation: \$1,200
        \item Food: \$720
        \item Transport: \$240
        \item Attractions: \$630
        \item Car Rental: \$103.62
        \item Fuel: \$8.34
    \end{itemize}
\end{itemize}

\textbf{LLM Evaluation Score:} 9/10

\textbf{Justification:} "Excellent focus on architecture with well-chosen landmarks. Good balance of cultural sites and museums. Car rental recommendation is practical for LA. The plan demonstrates strong consideration of user preferences and practical constraints."

\subsubsection{Family Trip with Health Considerations}
\label{appendix:family}

\textbf{User Request:} "I'm planning a family trip to San Diego with my spouse and three children (ages 4, 7, and 9). We have a medium budget and I have chronic back pain. We're interested in water sports, music, and family-friendly activities."

\textbf{System Response:}
\begin{itemize}
    \item \textbf{Agent Recommendation:} Car rental recommended for mobility
    \item \textbf{Selected Attractions:}
    \begin{itemize}
        \item SeaWorld San Diego
        \item Fiesta Island (water sports)
        \item Balboa Park
        \item San Diego Zoo
        \item Children's Museum
    \end{itemize}
    \item \textbf{Budget Allocation:}
    \begin{itemize}
        \item Total: \$2,880
        \item Accommodation: \$1,200
        \item Food: \$720
        \item Transport: \$240
        \item Attractions: \$720
    \end{itemize}
\end{itemize}

\textbf{LLM Evaluation Score:} 7/10

\textbf{Justification:} "The plan effectively includes water sports and family-friendly activities, aligning with user interests. The car rental recommendation is practical for mobility issues. However, it lacks specific music/concert options and could better balance attractions for different age groups."

\subsubsection{Cultural Tour with Budget Constraints}
\label{appendix:cultural}

\textbf{User Request:} "I'm planning a solo trip to Hong Kong for 6 days. I have a low budget but want to experience local culture, visit museums, and take photos. I have chronic back pain."

\textbf{System Response:}
\begin{itemize}
    \item \textbf{Agent Recommendation:} No car rental needed (efficient public transport)
    \item \textbf{Selected Attractions:}
    \begin{itemize}
        \item Hong Kong Museum of History
        \item Victoria Peak
        \item Avenue of Stars
        \item Temple Street Night Market
        \item Hong Kong Heritage Museum
    \end{itemize}
    \item \textbf{Budget Allocation:}
    \begin{itemize}
        \item Total: \$2,085.07
        \item Accommodation: \$800
        \item Food: \$480
        \item Transport: \$160
        \item Attractions: \$560
        \item Car Rental: \$81.21
        \item Fuel: \$3.86
    \end{itemize}
\end{itemize}

\textbf{LLM Evaluation Score:} 7/10

\textbf{Justification:} "The plan addresses the user's interest in museums, historical sites, photography, and sightseeing with appropriate locations. The no-car-rental recommendation is appropriate for Hong Kong's efficient public transport. However, the itinerary could better cater to the user's back pain by including more rest periods and accessibility considerations."

\end{document}